\documentclass[final,3p,times,twocolumn,authoryear]{elsarticle}
\usepackage{amssymb}
\usepackage{xspace}
\usepackage[utf8]{inputenc}
\usepackage{lineno}
\usepackage{url}
\usepackage{graphicx}
\usepackage{xcolor}
\usepackage{hyperref}
\usepackage{ecrc}
\usepackage{enumitem}

\makeatletter
\def\ps@pprintTitle{%
    \def\@evenhead{}%
    \let\@oddhead\@evenhead%
    \let\@oddfoot\@empty%
    \let\@oddfoot\@evenfoot}
\def\ps@headings{%
    \def\@oddhead{\parbox{\textwidth}{\itshape\footnotesize%
        \hfill\@runauth~/~AstroMLab-4\hfill{\rm \thepage}}}%
    \def\@evenhead{\parbox{\textwidth}{\itshape\footnotesize%
         {\rm \thepage}\hfil\@runauth~/~AstroMLab-4\hfil}}%
    \let\@evenfoot\@empty%
    \let\@evenfoot\@oddfoot}
\makeatother

\setlist[enumerate]{itemsep=0pt, topsep=0pt}
\usepackage{url}

\volume{00}
\firstpage{1}
\journalname{Astronomy and Computing}
\runauth{T. de Haan}
\jid{Astron. Comput.}
\jnltitlelogo{Astronomy and Computing}

\newcommand{\model}{AstroSage-Llama-3.1-70B\xspace}

\newcommand{\score}{$89.0\%$\xspace}

\newcommand{\modelsmall}{AstroSage-Llama-3.1-8B\xspace}

\journal{Astronomy $\&$ Computing}

\begin{document}

\begin{frontmatter}
\dochead{}

\title{AstroMLab 4: Benchmark-Topping Performance in Astronomy Q\&A with a 70B-Parameter Domain-Specialized Reasoning Model\\}

\author[1,2]{Tijmen de Haan}
\ead{tijmen.dehaan@gmail.com}
\author[3,4,5]{Yuan-Sen Ting}
\author[6]{Tirthankar Ghosal}
\author[7]{Tuan Dung Nguyen}
\author[8]{Alberto Accomazzi}
\author[6]{Emily Herron}
\author[6]{Vanessa Lama}
\author[9]{Rui Pan}
\author[10]{Azton Wells}
\author[10]{Nesar Ramachandra}

\address[1]{Institute of Particle and Nuclear Studies (IPNS), High Energy Accelerator Research Organization (KEK), Tsukuba, Ibaraki 305-0801, Japan}
\address[2]{International Center for Quantum-field Measurement Systems for Studies of the Universe and Particles (QUP-WPI), High Energy Accelerator Research Organization (KEK), Tsukuba, Ibaraki 305-0801, Japan}
\address[3]{Department of Astronomy, The Ohio State University, Columbus, OH, USA}
\address[4]{Center for Cosmology and AstroParticle Physics (CCAPP), The Ohio State University, Columbus, OH, USA}
\address[5]{Max-Planck-Institut für Astronomie, Heidelberg, Germany}
\address[6]{National Center for Computational Sciences, Oak Ridge National Laboratory, Oak Ridge, TN, USA}
\address[7]{Department of Computer and Information Science, University of Pennsylvania, Philadelphia, PA, USA}
\address[8]{Center for Astrophysics, Harvard \& Smithsonian, Cambridge, MA, USA}
\address[9]{Siebel School of Computing and Data Science, University of Illinois at Urbana-Champaign, Urbana-Champaign, IL, USA}
\address[10]{Computational Science Division, Argonne National Laboratory, Lemont, IL, USA}

\begin{abstract}
General-purpose large language models (LLMs), despite their broad capabilities, often struggle with specialized domain knowledge. This gap hinders their deployment as reliable research agents in demanding fields such as astronomy. Building on our prior work with \modelsmall, this study introduces \model, a 70-billion parameter domain-specialized natural-language AI assistant. It is designed for research and education across astronomy, astrophysics, space science, astroparticle physics, cosmology, and astronomical instrumentation. Developed from the Meta-Llama-3.1-70B foundation, \model underwent extensive continued pre-training (CPT) on a vast corpus of astronomical literature, followed by supervised fine-tuning (SFT) and model merging. We integrated reasoning chains into the SFT dataset, enabling \model to either answer the user query immediately, or first emit a human-readable thought process. Evaluated on a validated subset of 3,846 questions from the AstroMLab-1 benchmark \citep{tingAstroMLab1Who2024}---derived from literature withheld during training---\model achieves top-tier performance (89.0\%), matching GPT-5.2, Claude-4.5-Opus, and Gemini-3-Pro while being more cost-efficient. This work demonstrates that domain specialization, when applied to large-scale models, can enable them to outperform generalist counterparts in specialized knowledge areas like astronomy, thereby advancing the frontier of AI capabilities in the field.
\end{abstract}


\end{frontmatter}

\clearpage

\section{Introduction}
\label{introduction}
Astronomy and its related fields demand sophisticated tools that can process vast amounts of specialized knowledge. Large Language Models (LLMs) have emerged as promising assistants for this domain, offering capabilities as research collaborators, educational resources, and knowledge repositories. Domain-specialized models demonstrate cost-effectiveness in such contexts, as their parameters can be optimized for specific knowledge domains rather than distributed across the breadth of general internet content. Our previous work, \modelsmall \citep{de_haan_achieving_2025} (hereafter TdH25), established that an 8-billion parameter LLM, when trained on astronomical content, could match or exceed the performance of much larger general-purpose models on astronomical knowledge tasks \citep{tingAstroMLab1Who2024}. This finding highlighted the potential of domain specialization for creating efficient, high-performing AI assistants.

We now introduce \model, a 70-billion parameter language model that represents an advancement in specialized AI for astronomy. Our central research question asks whether domain specialization merely improves efficiency or can enable specialized models to outperform even the largest commercial alternatives. Following TdH25's approach with Meta-Llama-3.1-8B, we applied similar domain specialization techniques to the larger Meta-Llama-3.1-70B foundation. Beyond the increased parameter count, we implemented several key enhancements: expanded and refined datasets for both continued pre-training (CPT) and supervised fine-tuning (SFT); optimized learning hyperparameters based on public benchmarks and our own experimentation; and an explicit reasoning capability that enables step-by-step analytical processes before generating answers.

The core hypothesis driving this work is that a larger specialized model can elevate AI performance across astronomy, astrophysics, space science, astroparticle physics, cosmology, and astronomical instrumentation—collectively referred to as ``astronomy'' throughout this paper. While \modelsmall successfully matched larger models' performance, \model aims to surpass even advanced commercial alternatives. Beyond testing this hypothesis, we are making our trained model openly available to serve as a resource for researchers, educators, and students in the field, as well as non-scientists interested in astronomy.

This paper is structured as follows: Section \ref{sec:architecture_training} details the multi-stage training pipeline for \model, including improvements to our CPT and SFT corpora and procedures, and the model merging strategy for the final release. Section \ref{sec:features} discusses potential applications. Section \ref{sec:evaluation} evaluates \model against the updated \citet{tingAstroMLab1Who2024} benchmark, comparing it with other recent proprietary and open-weight models.

\section{Model Architecture and Training Methodology}
\label{sec:architecture_training}

\model is derived from the Meta-Llama-3.1-70B architecture \citep{dubeyLlama3Herd2024b}. This base model was selected for consistency with TdH25, which chose Meta-Llama-3.1-8B for its strong general capabilities and permissive licensing. The tokenizer from Meta-Llama-3.1-70B-Instruct was used without modification. Following the methodology of TdH25, our development process comprised three main stages: continued pre-training (CPT), supervised fine-tuning (SFT), and model merging.

\subsection{Continued Pre-training (CPT)}
\label{sec:pretraining}

The objective of CPT is to imbue the base model with domain-specific knowledge from astronomical literature. The CPT dataset for \model builds upon the comprehensive corpus developed in TdH25, which included approximately 250,000 arXiv preprints (astro-ph, gr-qc from 2007-2024), nearly 30,000 Wikipedia articles related to astronomy, and internet-available textbooks. The knowledge cutoff for the astro-ph papers included in the CPT dataset remains January 2024. 

This dataset was further enhanced:
\begin{itemize}
    \item We applied \textsc{ftfy} \citep{speer_ftfy_2019} for consistent Unicode text normalization and some rule-based repetition removal to correct OCR failures, supplementing the perplexity-based cleaning methods detailed in TdH25.
    \item To preserve general language understanding and mitigate catastrophic forgetting due to specialization, we incorporated a random selection of samples from the FineWeb dataset \citep{penedoRefinedWebDatasetFalcon2023} into each training epoch. This addition of previous pretraining tokens during CPT is sometimes known as ``replay''. The specific FineWeb samples were varied for each epoch, ensuring diverse exposure to general web text.
\end{itemize}

The CPT phase was conducted on the Oak Ridge Leadership Computing Facility (OLCF) Frontier supercomputer utilizing AMD Instinct MI250X GPUs. The Meta-Llama-3.1-70B base model underwent 2.5~epochs of CPT, a training process that utilized approximately 168,000~GPU-hours. Several learning hyperparameters were modified compared to TdH25. Firstly, the learning rate was scaled relative to that in \citet{dubeyLlama3Herd2024b} to maintain the same \textit{learning rate per token}. Secondly, the warm-up period was extended to 0.15~epochs. Finally, we increased the weight decay regularization term to 0.1, aligning with \citet{dubeyLlama3Herd2024b}.

\subsection{Supervised Fine-tuning (SFT)}
\label{sec:sft}

Following CPT, the model underwent SFT to develop its instruction-following and conversational capabilities, and to instill behaviors such as chain-of-thought and self-reflection (collectively termed ``reasoning'').

\begin{figure*}
\centering
\includegraphics[width=\linewidth]{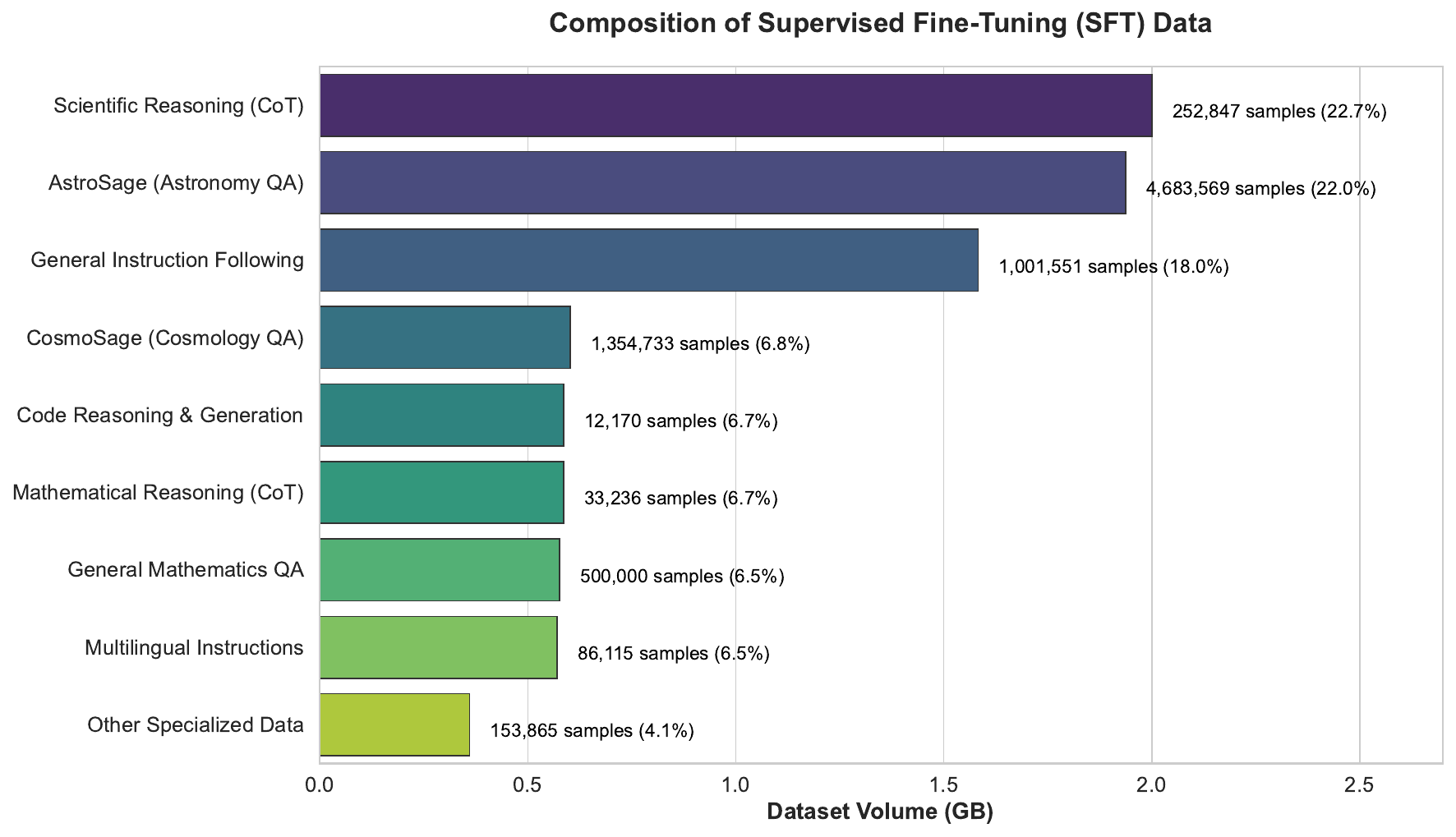}
\caption{Composition of the \model SFT training dataset. The combination of reasoning-focused datasets (22.7\%) with domain-specific astronomy Q\&A (22.0\%) reflects our strategy to develop a model that combines analytical thinking with specialized knowledge. General instruction-following data (OpenHermes 2.5) helps maintain versatility while preserving domain expertise. File sizes represent uncompressed UTF-8 encoded text including system and user prompts.}
\label{fig:sft_data}
\end{figure*}

Figure~\ref{fig:sft_data} illustrates the composition of the SFT dataset. Its largest component is NVIDIA's Llama-Nemotron-Post-Training-Dataset \citep{bercovich2025llamanemotronefficientreasoningmodels}, which was used to train the Llama-3\_1-Nemotron series. This series demonstrates strong performance on the LMArena \citep{chiangChatbotArenaOpen2024}, suggesting it is a strong dataset for eliciting reasoning and aligning with human preferences. The reasoning components cover science, code, mathematics, and general chat, providing a foundation for analytical thinking across different domains.

We also included the OpenHermes 2.5 dataset, which helps build general instruction-following capabilities and adherence to the system prompt.

To enhance domain expertise, we incorporated the custom domain-specific Q\&A datasets from both TdH25 and \citet{de_haan_cosmosage_2025}, which together comprise approximately 30\% of the training data. 

After combination, the dataset was deduplicated and shuffled. A loss mask was applied to train the model exclusively on assistant completions, excluding user queries and system prompts. The chat template adheres to the Llama-3.1 standard.

This diverse SFT mixture aims to produce a model that is knowledgeable in astronomy, has reasoning ability, and is good at following instructions.

The model was fine-tuned on this SFT dataset for 2 epochs. This phase consumed approximately 43,000 GPU-hours on the same MI250X GPU architecture. Hyperparameters mirrored those of the CPT stage, with the exception of weight decay, which was removed.

\subsection{Training Implementation}

The CPT and SFT stages were conducted on the Oak Ridge Leadership Computing Facility (OLCF) Frontier supercomputer using AMD Instinct MI250X GPUs. Our implementation employed the GPT-NeoX framework \citep{gpt_neox_library}, which we adapted for compatibility with the Llama-3.1 architecture. Training was distributed across 2,048 Graphics Compute Dies (GCDs) using a multi-dimensional parallelism strategy: tensor parallelism 8, pipeline parallelism 8, and data parallelism 32. As GPT-NeoX does not currently support DeepSpeed ZeRO stage 2/3 with pipeline parallelism, we used ZeRO stage 1 \citep{rasleyDeepSpeedSystemOptimizations2020} with activation checkpointing enabled. This configuration achieved a computational throughput of approximately 50 TFLOPS/s per GCD, consistent with performance metrics reported by \citet{dashOptimizingDistributedTraining2023}.

\begin{figure*}
\centering
\includegraphics[width=\linewidth]{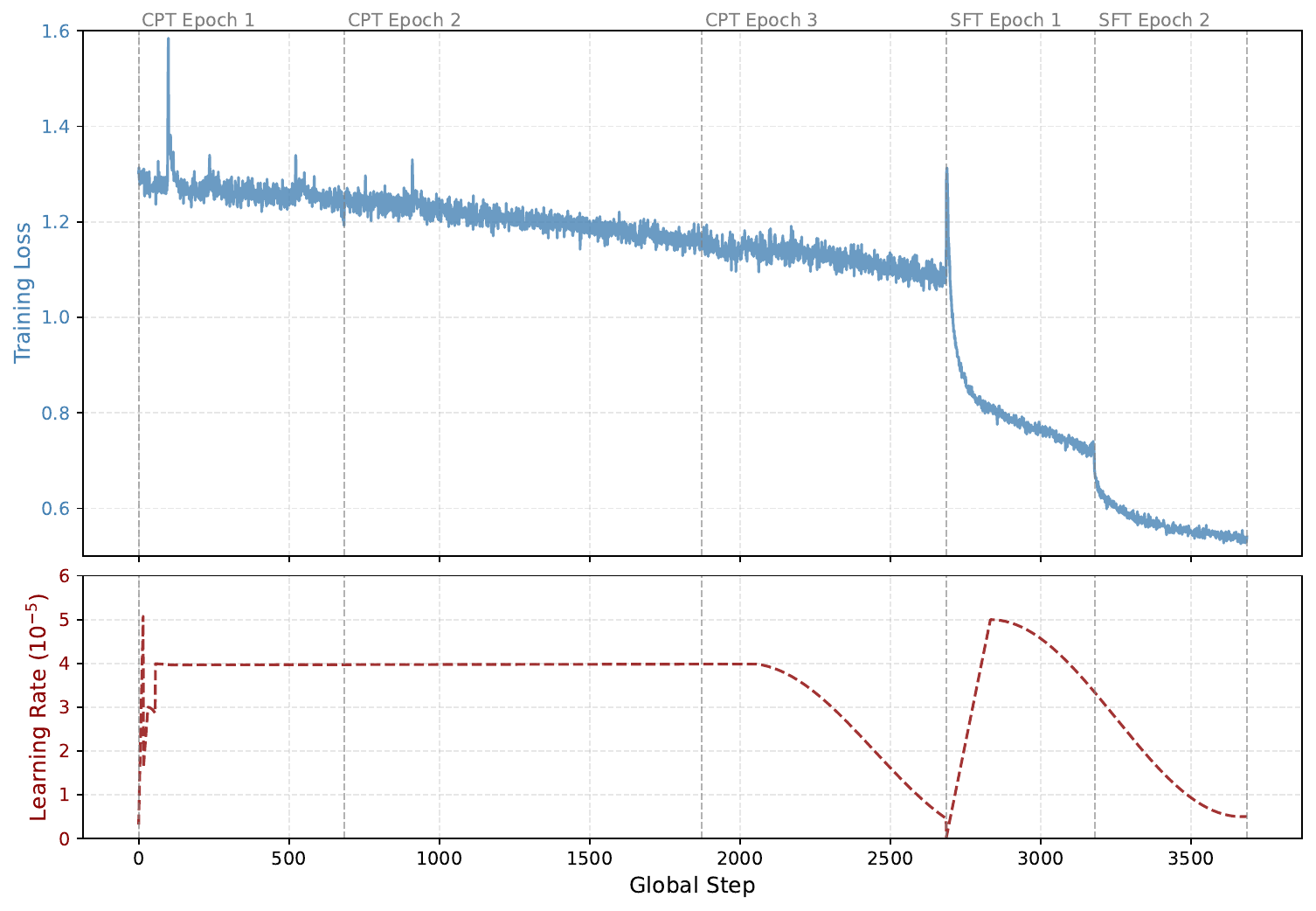}
\caption{Training dynamics for continued pre-training (CPT) and supervised fine-tuning (SFT). The top panel shows the training loss trajectory across 3~epochs of CPT followed by 2~epochs of SFT. Despite an early spike during CPT Epoch 1, the loss steadily decreases throughout CPT with minimal discontinuity at epoch boundaries, indicating effective learning without overfitting. The SFT phase exhibits a rapid initial decrease in loss, and shows discontinuous behavior at the epoch boundary, indicating the different statistical properties of the SFT data. The bottom panel shows the learning rate schedule, including the initial warm-up period, several manual adjustments during early CPT, the planned cosine decay during the final partial CPT epoch, and a separate warm-up and decay cycle for the SFT phase.}
\label{fig:loss_curve}
\end{figure*}

Figure~\ref{fig:loss_curve} illustrates the training dynamics during both CPT and SFT phases. Several patterns emerge from these curves. The CPT loss exhibits a consistent downward trajectory without discontinuities at epoch boundaries—a positive indicator that the model was still learning effectively from the dataset without overfitting. The SFT loss shows a rapid initial decrease and exhibits discontinuous behavior at the epoch boundary, reflecting the different statistical properties of the SFT data compared to CPT.

The SFT training loss initially exceeds the final CPT loss but decreases, dropping below 0.6. This indicates that, on average, the model assigns more than half its probability ($\left<P\right> = \exp(-\mathrm{loss})$) to the correct token—likely reflecting the more predictable nature of assistant responses compared to the CPT corpus. The model obtained after SFT is named AstroSage-70B-SFT.

\subsection{Model Merging}
\label{sec:merging}

To create the final, publicly released \model, we employed model merging (also known as parameter averaging) using the \texttt{mergekit} library \citep{goddardArceesMergeKitToolkit2024}. This technique allows us to combine the strengths of our specialized SFT model with the robust instruction-following capabilities of other popular fine-tuned Llama-3.1-70B variants.

The final mixture for \model was created using the DARE-TIES method \citep{yuLanguageModelsAre2024}. The AstroSage-70B-CPT model (our domain-specialized, pre-instruction-tuning model) was designated as the ``base model'' for the merge operation. The components and their contributions are:
\begin{itemize}
    \item 85\% AstroSage-70B-SFT (density 0.9)
    \item 15\% Meta-Llama-3.1-70B-Instruct (density 0.5)
\end{itemize}
The final model inherits the astronomical expertise from AstroSage-70B-SFT while incorporating the general instruction-following and conversational polish from Meta's Instruct model.

\section{Features and Reasoning Capability}
\label{sec:features}

\model is designed for a wide range of applications within the astronomical domain, offering capabilities similar to \modelsmall but enhanced by its larger scale and refined training. Potential applications include addressing factual queries (verifiable or low-risk), literature review, summarization, assisting with manuscript preparation and revision, brainstorming, hypothesis formulation, concept learning, programming support, and as a component in an agentic system.

A key feature of \model is its explicit reasoning capability. This aligns with recent advances in the broader LLM field, where explicit reasoning has emerged as a key development for enhancing model performance on complex tasks. The integration of reasoning mechanisms has become common in leading models, including OpenAI's o1 through o4 series, Anthropic's Claude models with ``thinking mode,'' DeepSeek-R1, and others. These developments demonstrate that exposing and structuring the internal reasoning process allows models to tackle complex problems systematically, resulting in improved accuracy and reliability.

Building on these industry-wide insights, \model implements explicit reasoning that can be activated at inference time by setting the system prompt to \texttt{detailed thinking on} and prefilling the assistant completion with \texttt{<think>}. When enabled, the model generates a step-by-step reasoning process before providing the final answer to the user's query. This is beneficial for complex astronomical problems requiring multi-step analysis. As the reasoning tokens are enclosed within \texttt{<think>} and \texttt{</think>} tags, they can easily be hidden from the end-user if desired.

\section{Evaluation}
\label{sec:evaluation}

\subsection{Benchmark Dataset and Methodology}
\label{sec:benchmark_dataset}

To evaluate \model, we utilize the AstroMLab-1 benchmark developed in \citet{tingAstroMLab1Who2024}. Unlike the original study, which used an internal version, we evaluate against the officially released version hosted on Hugging Face (\url{https://huggingface.co/datasets/AstroMLab/Astrobench_MCQ_v1_Public}). The original dataset comprises 4,425 multiple-choice questions derived from Annual Review of Astronomy and Astrophysics (ARAA) papers that were explicitly withheld from the AstroSage training corpus.

To ensure all questions are answerable based on general astronomical knowledge, we applied a cleaning procedure using Claude-3.5-Sonnet. Questions requiring access to specific paper content (e.g., particular numerical results or methodology choices unique to that study) were removed, and questions with unnecessary paper references were revised. This reduced the evaluation set to 3,846 validated questions.

For testing, we provided all models with minimal system prompts indicating they should act as astronomy experts. For models with explicit reasoning modes (e.g., o1, DeepSeek-R1, Gemini with ``thinking''), we enabled this feature. Reasoning modes did not improve scores for most models, likely because AstroMLab-1 questions primarily test knowledge recall rather than multi-step reasoning.

\subsection{Model Performance}

This evaluation significantly expands upon the initial benchmarking by \citet{tingAstroMLab1Who2024}. Using the refined, validated subset of AstroMLab-1, we assess the performance of 119 state-of-the-art models, superseding the preliminary results published in July 2024.

To quantify cost-efficiency, we define a \textit{value score} relative to the GPT-4 baseline (value = 0):
\begin{equation}
\mathrm{Value} = \frac{\mathrm{Score} - \mathrm{Score}_{\mathrm{GPT\textrm{-}4}}}{3.5} + \log_{10}\left(\frac{\mathrm{Cost}_{\mathrm{GPT\textrm{-}4}}}{\mathrm{Cost}}\right)
\end{equation}
Here, the denominator 3.5 represents the empirical performance slope observed in \citet{tingAstroMLab1Who2024}, where a tenfold increase in cost typically yields a 3.5 percentage point gain in accuracy. Intuitively, a Value Score of $+1$ signifies that a model delivers equivalent performance to the baseline while reducing inference costs by an order of magnitude ($10\times$). A score of $+5$ represents a $100,000\times$ improvement in cost-efficiency; for context, Gemini 3.0 Flash achieves a value of 4.98.

Table~\ref{tab:all_scores} presents the complete benchmark results organized by model family, reporting the score, model type (open-weight or proprietary), and value score.

\begin{table*}
\centering
\caption{Benchmark results on the AstroMLab-1 dataset (3,846 questions). Value scores are computed relative to GPT-4 (value = 0). Type: O = Open-weight, P = Proprietary.}
\label{tab:all_scores}
\footnotesize
\begin{tabular}{lr@{\,}cr|lr@{\,}cr|lr@{\,}cr}
\hline
Model & Score & T & Value & Model & Score & T & Value & Model & Score & T & Value \\
\hline
\multicolumn{12}{c}{\textit{AstroSage (Astronomy-Specialized)}} \\
\textbf{\modelsmall} & \textbf{81.9} & \textbf{O} & \textbf{+4.12} & \textbf{\model} & \textbf{89.0} & \textbf{O} & \textbf{+5.55} & & & & \\
\hline
\multicolumn{4}{l|}{\textit{OpenAI/GPT}} & \multicolumn{4}{l|}{\textit{Alibaba/Qwen}} & \multicolumn{4}{l}{\textit{Zhipu/GLM}} \\
GPT-3.5 & 72.7 & P & +0.21 & Qwen 7B & 59.2 & O & $-$2.12 & GLM-3 Turbo & 66.8 & P & $-$0.64 \\
GPT-4 & 77.8 & P & 0.00 & Qwen 1.5 7B & 65.9 & O & $-$0.22 & GLM-4 Flash & 69.5 & P & +1.14 \\
GPT-4o-mini & 78.9 & P & +2.39 & Qwen 1.5 14B & 70.1 & O & +0.78 & GLM-4 AirX & 74.9 & P & +0.68 \\
GPT-4o & 85.1 & P & +2.94 & Qwen 1.5 32B & 75.4 & O & +1.98 & GLM-4 Air & 75.2 & P & +1.79 \\
GPT-4.1-nano & 80.8 & P & +3.11 & Qwen 1.5 110B & 75.4 & O & +1.50 & GLM-4 0520 & 77.6 & P & +0.47 \\
GPT-4.1-mini & 86.7 & P & +4.19 & Qwen 2 7B & 69.9 & O & +0.92 & GLM-4 Plus & 80.5 & P & +2.43 \\
GPT-4.1 & 86.6 & P & +3.47 & Qwen 2 57B & 73.9 & O & +1.26 & GLM-4 9B & 72.3 & O & +1.40 \\
GPT-5-nano & 83.1 & P & +3.82 & Qwen 2 72B & 80.5 & O & +3.03 & GLM-4 32B & 75.6 & O & +1.96 \\
GPT-5-mini & 86.7 & P & +4.15 & Qwen 2.5 7B & 72.5 & O & +1.31 & GLM-4.5 Air & 82.2 & P & +3.22 \\
GPT-5 & 88.3 & P & +3.92 & Qwen 2.5 14B & 73.1 & O & +1.48 & GLM-4.5 & 85.3 & O & +3.62 \\
GPT-5.2 & 89.4 & P & +4.09 & Qwen 2.5 32B & 73.3 & O & +1.25 & GLM-4.6 & 85.1 & O & +3.46 \\
o1-mini & 82.6 & P & +2.17 & Qwen 2.5 72B & 81.0 & O & +3.00 & GLM-4.7 & 84.9 & O & +3.54 \\
o1 & 86.2 & P & +2.49 & Qwen 2.5 Max & 83.5 & P & +2.69 & \multicolumn{4}{l}{\textit{Google/Gemini}} \\
o3-mini & 86.2 & P & +3.64 & Qwen 3 14B & 79.0 & O & +2.89 & Gemini 1.0 Pro & 73.3 & P & +0.37 \\
o3 & 87.2 & P & +3.64 & Qwen 3 30B-A3B & 81.2 & O & +3.50 & Gemini 1.5 Flash & 78.8 & P & +2.04 \\
o4-mini & 86.2 & P & +3.63 & Qwen 3 32B & 82.1 & O & +3.69 & Gemini 1.5 Pro & 80.5 & P & +1.59 \\
gpt-oss-20b & 84.0 & O & +4.67 & Qwen 3 80B & 83.2 & O & +3.85 & Gemini 2.0 Flash & 83.2 & P & +3.82 \\
gpt-oss-120b & 85.7 & O & +4.83 & Qwen 3 235B & 86.3 & O & +4.53 & Gemini 2.0 Pro & 85.9 & P & +3.26 \\
\multicolumn{4}{l|}{\textit{Meta/Llama}} & QwQ 32B & 81.1 & O & +3.18 & Gemini 2.5 Flash & 83.9 & P & +3.27 \\
Llama 2 7B & 51.1 & O & $-$4.45 & \multicolumn{4}{l|}{\textit{Anthropic/Claude}} & Gemini 2.5 Pro & 86.9 & P & +3.52 \\
Llama 2 70B & 72.6 & O & +0.89 & Claude 2 & 77.8 & P & +0.47 & Gemini 3.0 Flash & 90.3 & P & +4.98 \\
Llama 3 8B & 75.2 & O & +2.22 & Claude 3 Haiku & 80.6 & P & +2.59 & Gemini 3.0 Pro & 88.9 & P & +3.99 \\
Llama 3 70B & 82.8 & O & +3.78 & Claude 3 Sonnet & 79.8 & P & +1.29 & \multicolumn{4}{l}{\textit{Google/Gemma}} \\
Llama 3.1 8B & 75.7 & O & +2.36 & Claude 3 Opus & 85.0 & P & +2.07 & Gemma 2B & 44.9 & O & $-$6.05 \\
Llama 3.1 70B & 82.5 & O & +3.60 & Claude 3.5 Haiku & 78.0 & P & +1.33 & Gemma 7B & 57.5 & O & $-$2.60 \\
Llama 3.1 405B & 85.8 & O & +3.54 & Claude 3.5 Sonnet & 86.4 & P & +2.88 & Gemma 2 2B & 60.5 & O & $-$1.60 \\
Llama 3.2 1B & 47.3 & O & $-$5.34 & Claude 3.7 Sonnet & 86.3 & P & +3.13 & Gemma 2 9B & 73.7 & O & +1.72 \\
Llama 3.2 11B & 76.7 & O & +2.50 & Claude 4 Sonnet & 86.9 & P & +3.30 & Gemma 2 27B & 77.5 & O & +1.75 \\
Llama 3.2 90B & 82.9 & O & +3.51 & Claude 4 Opus & 87.8 & P & +2.87 & Gemma 3 4B & 69.1 & O & +0.56 \\
Llama 3.3 70B & 82.6 & O & +3.72 & Claude 4.5 Haiku & 83.3 & P & +2.75 & Gemma 3 12B & 76.2 & O & +2.39 \\
Llama 4 Scout & 83.6 & O & +4.13 & Claude 4.5 Sonnet & 87.3 & P & +3.42 & Gemma 3 27B & 78.7 & O & +2.91 \\
Llama 4 Maverick & 85.5 & O & +4.38 & Claude 4.5 Opus & 89.0 & P & +3.68 & \multicolumn{4}{l}{\textit{Microsoft/Phi}} \\
\multicolumn{4}{l|}{\textit{DeepSeek}} & \multicolumn{4}{l|}{\textit{Mistral AI}} & Phi-2 & 66.3 & O & +0.08 \\
DeepSeek 67B & 65.3 & O & $-$0.61 & Mistral 7B & 66.1 & O & $-$0.28 & Phi-3 mini & 73.6 & O & +1.99 \\
DeepSeek V2 & 75.9 & O & +1.80 & Mixtral 8x7B & 76.7 & O & +1.61 & Phi-3 medium & 77.7 & O & +2.74 \\
DeepSeek V2.5 & 78.7 & O & +2.77 & Mixtral 8x22B & 80.4 & O & +1.79 & Phi-3.5 mini & 74.6 & O & +2.28 \\
DeepSeek V3 & 85.3 & O & +3.92 & Mistral Small 3.1 & 81.3 & O & +3.82 & Phi-4 mini & 71.4 & O & +1.36 \\
DeepSeek V3.1 & 85.8 & O & +4.30 & Mistral Medium 3.1 & 84.0 & P & +3.35 & Phi-4 & 82.2 & O & +3.91 \\
DeepSeek V3.2 & 84.9 & O & +4.17 & Mistral Large 1 & 79.3 & P & +2.09 & \multicolumn{4}{l}{\textit{xAI/Grok}} \\
DeepSeek R1 & 88.2 & O & +4.42 & Mistral Large 2 & 83.1 & P & +2.57 & Grok Beta & 82.0 & P & +1.91 \\
\multicolumn{4}{l|}{\textit{Moonshot/Kimi}} & Mistral Large 3 & 85.0 & O & +3.73 & Grok 2 & 84.0 & P & +2.67 \\
Moonshot v1 & 74.6 & P & +0.52 & \multicolumn{4}{l|}{\textit{MiniMax}} & Grok 3 Mini & 85.6 & P & +4.30 \\
Kimi K2 & 87.5 & O & +4.23 & MiniMax-01 & 83.4 & P & +2.55 & Grok 4 Fast & 85.6 & P & +4.35 \\
& & & & MiniMax-M2 & 84.5 & P & +3.69 & & & & \\
& & & & MiniMax-M2.1 & 83.8 & O & +3.90 & & & & \\
\hline
\end{tabular}
\end{table*}

\begin{figure*}
\centering
\includegraphics[width=\linewidth]{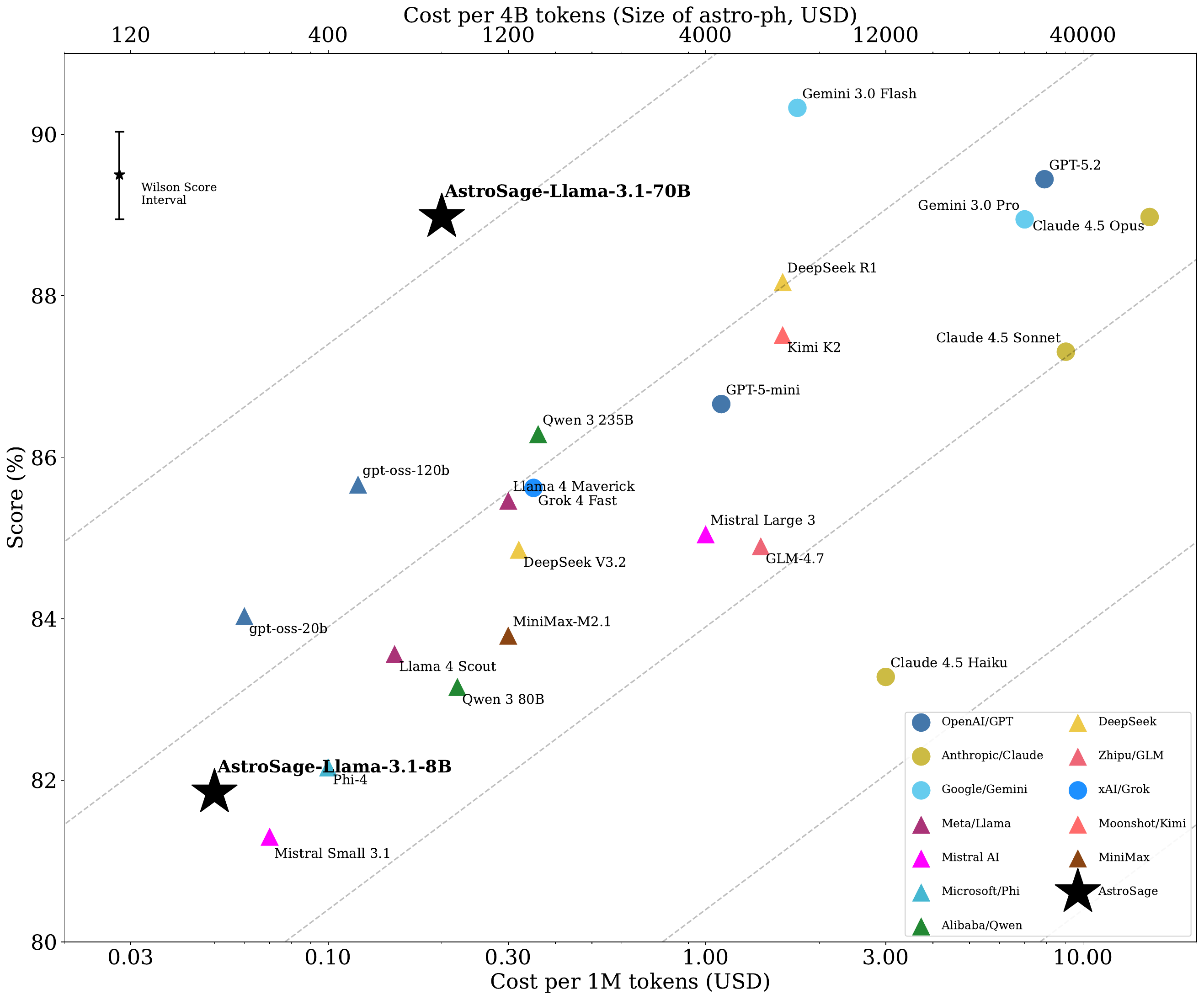}
\caption{Performance comparison on the AstroMLab-1 benchmark. The x-axis shows cost per 0.1M tokens (USD) on a logarithmic scale; the y-axis shows accuracy percentage. Open-weight models are shown as triangles; proprietary models as circles. The diagonal dashed lines represent cost-efficiency trade-offs where a tenfold increase in cost corresponds to approximately 3.5 percentage points improvement. The Wilson Score interval shows uncertainty due to finite sample size of questions.}
\label{fig:benchmark_results}
\end{figure*}

On this benchmark, \model achieves a score of \score, placing it among the top models alongside GPT-5.2 (89.4\%), Claude~4.5~Opus (89.0\%), and Gemini~3.0~Pro (88.9\%)---all released in late 2025, months after \model was finalized in May 2025. Gemini~3.0~Flash (90.3\%) is the only model to surpass \model in absolute score; despite being a ``Flash'' variant, it outperforms even its own Pro version. However, Gemini~3.0~Flash achieves a lower value score (+4.98 vs.\ +5.55) due to higher inference cost---in practical terms, processing the entire astro-ph archive would cost approximately \$10,000 with Gemini~3.0~Flash versus \$1,000 with \model. For context, professional astronomers score around 67\% on this benchmark \citep{de_haan_achieving_2025}.

Figure~\ref{fig:benchmark_results} illustrates the cost-performance landscape. For cost analysis, we averaged input and output token costs as listed at \url{https://llmtokencost.com}, cross-verified with \url{https://artificialanalysis.ai}. For AstroSage models, we assume the same inference cost as Llama-3.1 models of equivalent size, as they share identical architectures. While API pricing changes over time, such variations do not qualitatively alter the comparative landscape. As identified in \citet{tingAstroMLab1Who2024}, leading models cluster along diagonal lines where a tenfold cost increase corresponds to roughly 3.5 percentage points in accuracy---a relationship that persists even as the frontier has shifted upward since summer 2024.

\subsection{Temporal Evolution}
\label{sec:temporal}

\begin{figure*}
\centering
\includegraphics[width=\linewidth]{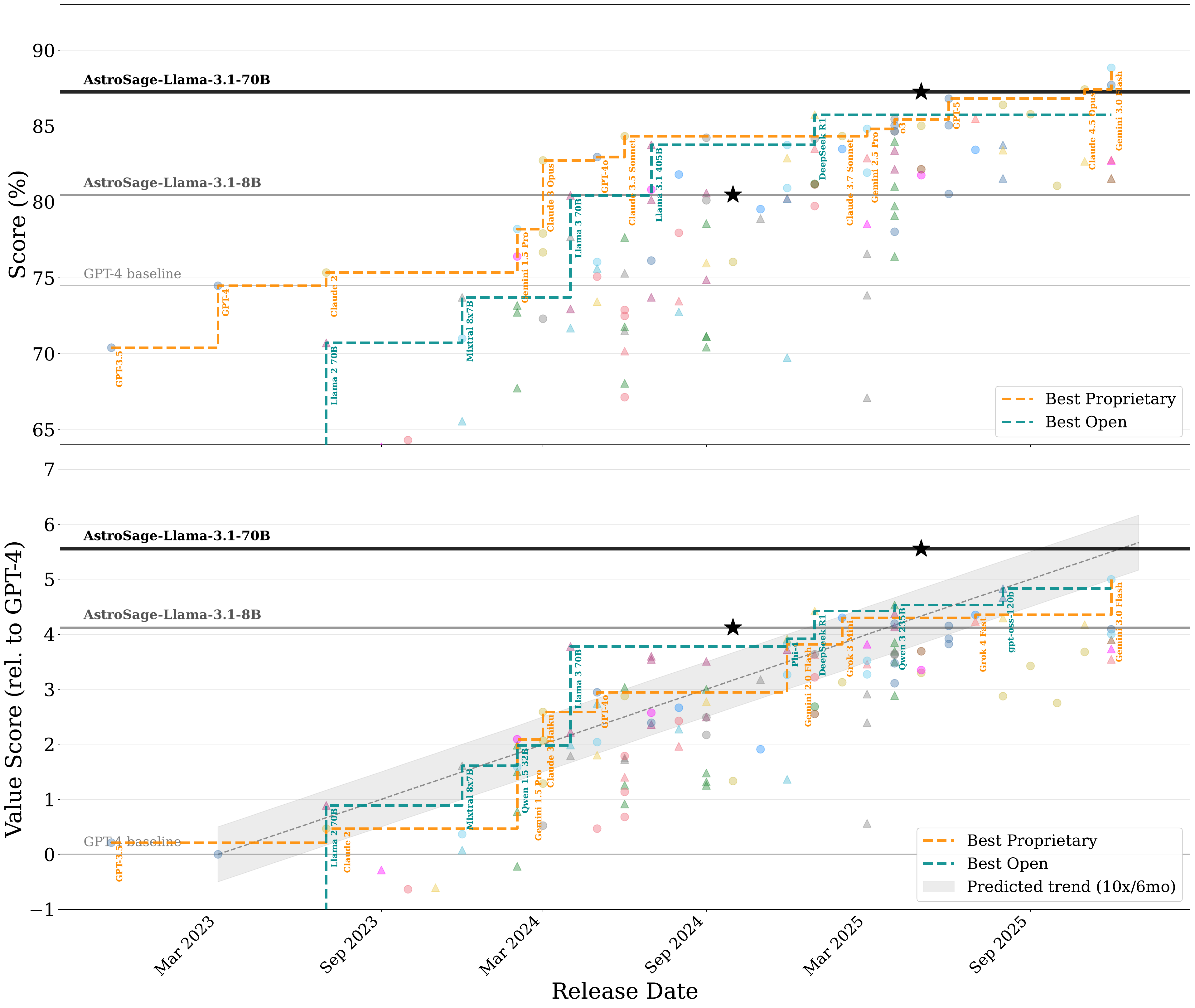}
\caption{Evolution of LLM performance over time. Colors and symbols follow Figure~\ref{fig:benchmark_results}. \textbf{Top panel:} Absolute score versus release date, with step functions showing the best open-weight (teal) and proprietary (orange) models. \textbf{Bottom panel:} Value score versus time. The shaded region shows the predicted trend of +1 value (10$\times$ cost-efficiency) per six months, anchored at GPT-4. Horizontal lines indicate AstroSage performance levels with stars at release dates.}
\label{fig:time_evolution}
\end{figure*}

Figure~\ref{fig:time_evolution} presents the temporal evolution of model performance. A key prediction from \citet{tingAstroMLab1Who2024} was that cost-efficiency would improve by 10$\times$ every six months. The shaded region in the bottom panel, anchored at GPT-4's release and projecting this trend forward, tests this prediction. After more than two years, this projection continues to track the frontier.

The gap between open-weight and proprietary frontiers has narrowed. While proprietary models led through early 2024, open alternatives such as DeepSeek-R1, Qwen-3, and the GPT-OSS series now approach parity. This is relevant for astronomy, where open models enable deployment on institutional infrastructure and processing of sensitive data.

The key result is that \model, as a 70-billion parameter open-weight model, achieves scores matching the top proprietary flagships (GPT-5.2, Claude~4.5~Opus, Gemini~3.0~Pro) while being 35--75$\times$ cheaper to run. This cost advantage drives its higher value score.  Both AstroSage (8B and 70B) models improve 6.2-6.5 points compared to their base models (Llama-3.1-8B and Llama-3.1-70B), representing $\sim$100$\times$ cost-efficiency improvement through domain specialization---an advantage that general scaling alone cannot match.

An emerging trend is that performance gaps between flagship models and their smaller variants have widened. For example, Claude~4.5~Opus (89.0\%) outperforms Claude~4.5~Haiku (83.3\%) by 5.7 percentage points, and GPT-5.2 (89.4\%) exceeds GPT-5-nano (83.1\%) by 6.3 points. This widening gap reinforces the value of domain-specific training for deploying cost-effective models.

We acknowledge that current astronomy benchmarks offer limited evaluation of problem-solving capabilities requiring deep reasoning. Future work should develop more complex reasoning-based astronomy benchmarks to better assess the full potential of reasoning-enhanced models in this domain.

\section{Availability}
\label{sec:availability}

To encourage adoption, further research, and community-driven improvements, \model is made freely available under the Llama 3.1 Community License. The full model weights can be accessed and downloaded from our project repository on Hugging Face: \url{https://huggingface.co/AstroMLab/AstroSage-70B-20251009}.

We hope that by providing open access to \model, we can accelerate the development and application of advanced AI tools within the astronomical community.

\section{Discussion and Conclusion}
\label{sec:discussion}

The development of \model advances specialized language models for astronomy. Building on the foundation established by \modelsmall, this 70-billion parameter model incorporates a more powerful base architecture, enriched training datasets, refined training methodologies, and explicit reasoning capabilities. On the AstroMLab-1 benchmark, \model achieved a score of \score, placing it among the top-performing models alongside advanced proprietary offerings, while being more cost-efficient.

Our results support the central hypothesis of this work: domain specialization, when scaled to larger models, can enable specialized systems to compete with advanced general-purpose commercial models within their domain of expertise. However, it is crucial to note that this strong performance on astronomy-specific tasks does not necessarily translate to general capabilities. In tasks requiring complex mathematical reasoning or general problem-solving, more generalized models likely maintain advantages. Nevertheless, through internal testing and preliminary expert evaluations, we have observed that \model demonstrates improved domain knowledge recall and the ability to make connections between astronomical concepts compared to general-purpose models. While these qualitative assessments are promising, we acknowledge that more rigorous, large-scale arena-type testing would be necessary to fully validate these observations.

The practical implications of these findings for astronomy are still emerging. While \model at 70B parameters represents a computational investment, it may prove valuable for tasks that require both scalability and domain expertise, such as processing and summarizing large volumes of astronomical literature. The open-weight nature of \model also enables offline processing of proprietary data from institutions like NASA and NSF, addressing a practical consideration for many research applications.

Future development must address two emerging frontiers. First, the field requires benchmarks that transcend simple knowledge retrieval. To fully assess models like \model, we must develop evaluation suites that test multi-step reasoning, hypothesis formulation, and data interpretation in realistic research scenarios. Second, the potential of this model will be best realized through integration with astronomical software pipelines. Moving beyond a chat interface to an agentic framework---where the model can query databases (e.g., SIMBAD, ADS) or execute analysis code---represents the next leap in AI-assisted discovery.

\model advances the integration of AI assistants into astronomical research and education. By making such specialized tools openly available, we aim to democratize access to cutting-edge AI and accelerate scientific discovery. As astronomical data continues to grow in volume and complexity, tools like \model will be valuable for extracting insights, generating hypotheses, and accelerating the pace of scientific discovery.

We welcome feedback from the astronomical community and invite collaboration to further refine and expand upon this work. The open-source nature of \model provides a foundation for community-driven improvements and adaptations to specific research needs within astronomy and related fields.

\section*{Acknowledgements}

This research used resources of the Oak Ridge Leadership Computing Facility (OLCF), which is a DOE Office of Science User Facility at the Oak Ridge National Laboratory supported by the U.S. Department of Energy under Contract No. DE-AC05-00OR22725 and support from Microsoft's Accelerating Foundation Models Research (AFMR) program. TdH was supported by World Premier International Research Center Initiative (WPI), MEXT, Japan. YST is supported by the National Science Foundation under Grant No. 2406729 and a Humboldt Research Award from the Alexander von Humboldt Foundation. Work at Argonne National Laboratory is supported by UChicago Argonne LLC, Operator of Argonne National Laboratory. Argonne, a U.S. Department of Energy Office of Science Laboratory, is operated under Contract No. DE-AC02-06CH11357. The ongoing public deployment of \model is supported by ACCESS allocation PHY240259. A special thanks to Sajal Dash for suggesting the use of the GPT-NeoX LLM Training framework.

% Redefine the bibliography heading to be "REFERENCES" in uppercase
\makeatletter
\renewcommand\bibsection{\section*{\MakeUppercase{\refname}}}
\makeatother

\bibliography{astrosage}

\end{document}